\documentclass[ a4paper, 12pt ]
  {article}

\usepackage[dvips]{graphicx}

\newcommand{\bfk}{\mbox{$\mathbf k$}}
\newcommand{\NP}[1]{{\it Nucl.\ Phys.}\ {\bf #1}}
\newcommand{\ZP}[1]{{\it Z.\ Phys.}\ {\bf #1}}
\newcommand{\PL}[1]{{\it Phys.\ Lett.}\ {\bf #1}}
\newcommand{\PR}[1]{{\it Phys.\ Rev.}\ {\bf #1}}

\newcommand{\IJMP}[1]{{\it Int.\ J.\ Mod.\ Phys.}\ {\bf #1}}
\begin{document}

\begin{center}
\bf{Single spin asymmetries, unpolarized cross sections and \\
the role of partonic transverse momentum}\footnote{ 
Talk delivered by U.~D'Alesio at the 
``15th International Spin Physics Symposium'', SPIN2002, September 9-14, 2002, 
Brookhaven National Laboratory, Upton (NY), USA. }

\vspace*{0.6cm}

{\sf U.~D'Alesio, F.~Murgia}
\vskip 0.5cm

{\it Istituto Nazionale di Fisica Nucleare, 
Sezione di Cagliari \\
and Dipartimento di Fisica, Universit\`a di Cagliari \\
C.P. 170, I-09042 Monserrato (CA), Italy} \\
\end{center}

\vspace*{0.5cm}

\begin{abstract}
Partonic intrinsic transverse momentum can be essential for the
explanation of large single spin asymmetries in hadronic reactions
in the framework of perturbative QCD.
The status of an ongoing program investigating in a consistent
way the role of intrinsic transverse momentum both in unpolarized and
polarized processes is discussed.
We compute inclusive cross sections for hadron and photon production in
hadronic collisions and for Drell-Yan processes; the results are
compared
with available experimental data in several different kinematical
situations.
\end{abstract}

\vspace*{0.5cm}

In the last years a lot of experimental and theoretical activity has
been devoted to the study of transverse single spin asymmetries (SSA)
in hadronic collisions and in semi-inclusive DIS.
In fact, perturbative QCD (pQCD) with ordinary
collinear partonic kinematics leads to negligible values
for these asymmetries, as soon as the relevant scale of the process under
consideration becomes large. There are however several experimental results
which seem to contradict this expectation; two well known examples are:
the large transverse $\Lambda$ polarization measured in unpolarized
hadronic collisions; the large SSA observed in the process
$p^\uparrow p\to \pi\,X$. A possible way out from this situation comes from
extending the collinear pQCD formalism with the inclusion of spin and
partonic intrinsic transverse momentum, $\bfk_\perp$, effects.
This leads to
the introduction of a new class of spin and $\bfk_\perp$ dependent
partonic distribution (PDF) and fragmentation (FF) functions, describing
fundamental properties of hadron structure \cite{anse}.

The role of $\bfk_\perp$ effects in inclusive hadronic reactions
has been extensively studied also in the calculation of
unpolarized cross sections. 
It has been shown that, particularly at moderately large $p_{_T}$ (which is the
region where SSA are measured to be large) these effects can be relevant and
may help in improving the agreement between experimental results and pQCD
(at LO and NLO) calculations, which often underestimate the data
\cite{ww}. 

Based on these considerations, in this contribution we present a preliminary
account of an ongoing program which aims to describe consistently both
polarized and unpolarized cross sections (and SSA) for
inclusive particle production in hadronic collisions at
large energies and moderately large $p_{_T}$, using LO pQCD with
the inclusion of intrinsic transverse momentum effects.
Our main goal is not to fit the cross sections
as well as possible (including NLO contributions, etc.), but rather to show
that in our LO approach they are reproduced up to an overall factor
of 2-3, compatible with expected NLO K-factors and scale dependences,
which reasonably cancel out in SSA and are then out of our present interest.

In a pQCD approach at LO and leading twist
with inclusion of spin and $\bfk_\perp$ effects, the
unpolarized cross section for the inclusive process $A\,B\to C\,X$
reads 
\begin{equation}
d\sigma \!\propto\! \sum_{a,b,c} \hat f_{a/A}(x_a,\bfk_{\perp\,a}) \otimes
\hat f_{b/B}(x_b, \bfk_{\perp\,b}) \otimes
d\hat\sigma^{ab \to c \dots}(x_a, x_b, \bfk_{\perp\,a},
 \bfk_{\perp\,b}) \otimes \hat D_{C/c}(z, \bfk_{\perp\,C})\,,
\label{ucr}
\end{equation}
\noindent with obvious notations.
A similar expression holds for the numerator of a transverse
SSA ($\propto
d\,\Delta^{\!N}\!\sigma/d\sigma$), substituting for the polarized particle
involved the corresponding unpolarized PDF (or FF)
with the appropriate polarized one, $\Delta^{\!N}\!f$ or
$\Delta^{\!N}\!D$.
At leading twist there are four new spin and $\bfk_\perp$ dependent
functions to take into account:
\begin{eqnarray}
\!\!\!\!\!\!\!\!\!\!\!\!\!
\Delta^{\!N}\!f_{q/p^\uparrow}  \!\!\! &\equiv& \!\!\!
\hat f_{q/p^\uparrow}(x, \bfk_{\perp})-\hat f_{q/p^\downarrow}
(x, \bfk_{\perp})\>;\>\>
\Delta^{\!N}\!f_{q^\uparrow/p} \equiv
\hat f_{q^\uparrow/p}(x, \bfk_{\perp})-\hat f_{q^\downarrow/p}
(x, \bfk_{\perp})\>; \label{delf2} \\
\!\!\!\!\!\!\!\!\!\!\!\!\!
\Delta^{\!N}\!D_{h/q^\uparrow} \!\!\!\! &\equiv& \!\!\!\!
\hat D_{h/q^\uparrow}(z, \bfk_{\perp}) -
\hat D_{h/q^\downarrow}(z, \bfk_{\perp})\>;\>\>
\Delta^{\!N}\!D_{h^\uparrow/q} \equiv
\hat D_{h^\uparrow/q}(z, \bfk_{\perp}) -
\hat D_{h^\downarrow/q}(z, \bfk_{\perp})\,,
\label{deld2}
\end{eqnarray}
\noindent two in the PDF, Eq. (\ref{delf2}),
and two in the FF, Eq. (\ref{deld2});
the first functions in Eq.s (\ref{delf2}),(\ref{deld2}) are
respectively the Sivers and the Collins function.
The second ones are respectively the function introduced by Boer
\cite{boer} and the so-called ``polarizing'' FF \cite{abdm}.

The unpolarized PDF and FF are given in a
simple factorized form, and the $\bfk_\perp$ dependent part
is usually taken to have a Gaussian shape:
\begin{equation}
\hat f_{a/A}(x,\bfk_{\perp a}) = f_{a/A}(x)\,\frac{\beta^2}{\pi}\>
e^{-\beta^2\,k_{\perp a}^{\,2}}\>;
\qquad\qquad
\hat D_q^h(z,\bfk_{\perp h}) = D_q^h(z)\,\frac{\beta'^2}{\pi}\>
e^{-\beta'^2\,k_{\perp h}^{\,2}}\>,
\label{gk}
\end{equation}
\noindent where the parameter $\beta$ ($\beta'$) is related to the
average partonic (hadronic) $k_{\perp}$ by the simple relation  
$1/\beta(\beta')=\langle\,k_{\perp a(h)}^2\,\rangle^{1/2}$.
Similar expressions are adopted for the polarized PDF and FF of
Eq.s (\ref{delf2}),(\ref{deld2}).

Using this approach, we have studied the unpolarized cross section for
several hadronic processes, analyzing a large sample of available data
in different kinematical situations:
{\it i)} The Drell-Yan process $p\,p\to \ell^+\ell^-\,X$;
{\it ii)} Prompt photon production in $p\,p\to\gamma\,X$;
{\it iii)} Inclusive pion production in $p\,p\to\pi\,X$.
We find that an overall good reproduction of the corresponding
unpolarized cross sections is possible (within the limits indicated
above) by choosing, depending on the kinematical situation
considered,
$\beta=1.0-1.25$ (GeV/$c)^{-1}$ (that is,
$\langle\,k_\perp^2\,\rangle^{1/2}=0.8-1.0$ GeV/$c$).
The choice of $\beta$ is related to the set
of $x$ dependent PDF utilized; throughout this paper we use the
GRV94 set \cite{grv94}. The optimal choice of $\beta'$ in case
{\it iii)} (pion production) is commented in the following. 
We limit ourself to present few indicative results and comments
regarding the processes considered and the SSA in pion production.
A full account of this analysis will be presented elsewhere \cite{damu}.

\noindent {\it i)}  At LO and within collinear partonic configuration 
the final lepton pair produced in Drell-Yan processes
cannot have any transverse momentum, $q_{_T}$, with respect to the
colliding beams. Experimental data show however that the lepton pair
has a well defined $q_{_T}$ spectrum.
As an example, in Fig. 1a we show estimates of the invariant cross section
at $E$ = 400 GeV as a function of $q_{_T}$,
for several different invariant mass bins (in GeV) at fixed rapidity
$y$ = 0.03, and using $\beta=1.11$ (GeV$/c)^{-1}$;
data are from \cite{ito}.
Theoretical curves are arbitrarily raised by a factor $K_{\rm fac}=1.6$,
which could be well accommodated by NLO K-factors and scale dependences,
an issue that as said above we do not address here.
Notice how data are well reproduced by a
Gaussian dependence up to $q_{_T} =2-2.5$ GeV/$c$; larger $q_{_T}$ data
show a power-law decrease well explained by pQCD corrections.

\noindent {\it ii)}
Several data for the unpolarized cross section are available in the case
of prompt photon production, mainly at central rapidities and moderately
large $p_{_T}$. As an example, in Fig. 1b
we show estimates of the invariant cross section for the process
$p\,p\to\gamma\,X$ at $E$ = 280 GeV for two different values of
$p_{_T}$ vs. $x_{_F}$, with $\bfk_{\perp}$ effects (thick lines)
and without them (thin lines), using $K_{\rm fac}=1$ and
$\beta=1.25$ (GeV$/c)^{-1}$; data are from \cite{bone}.

\begin{figure}
\includegraphics[angle=-90,width = 7.7cm]{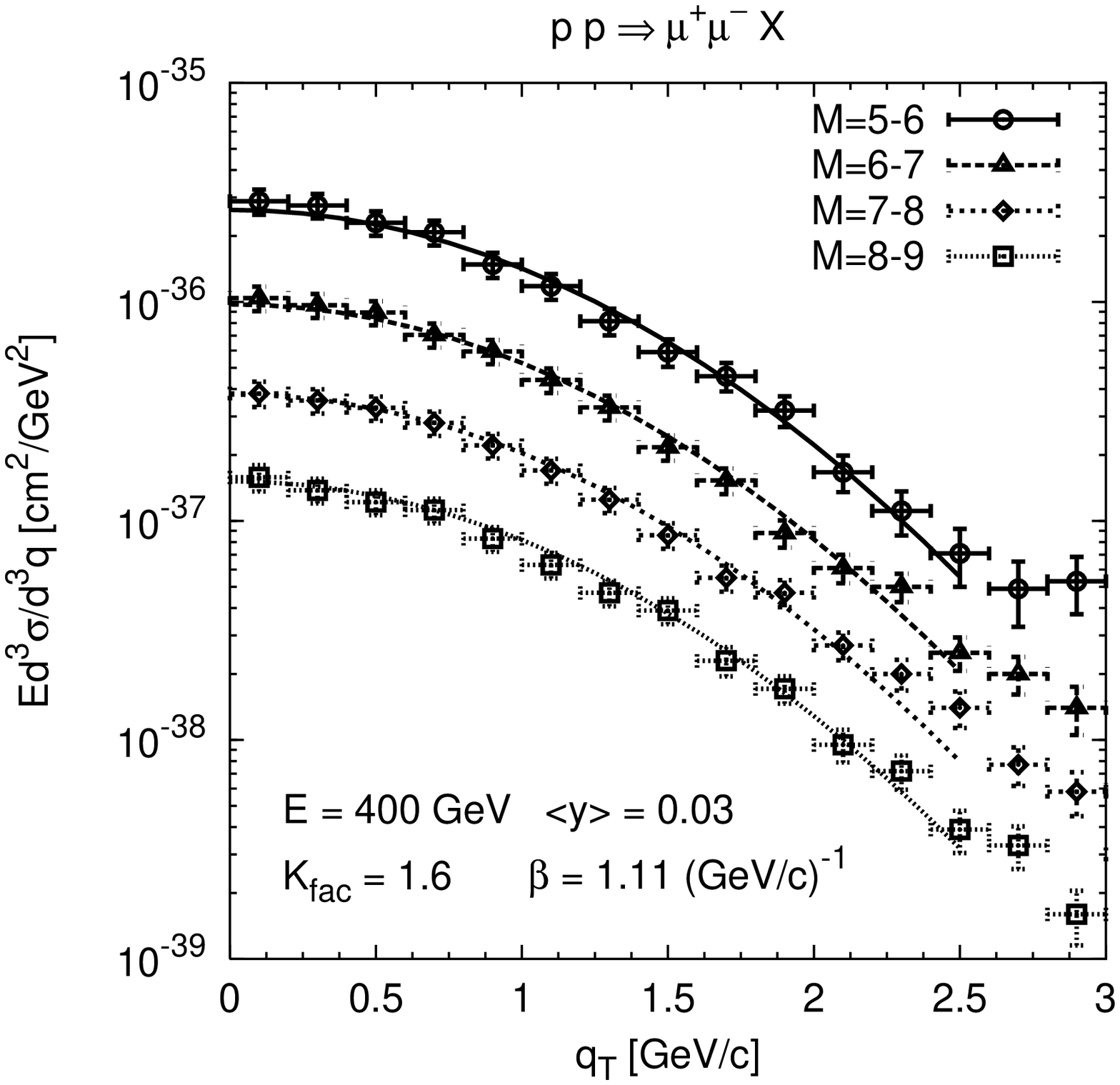}
\hspace*{-10pt}
\includegraphics[angle=-90,width = 7.7cm]{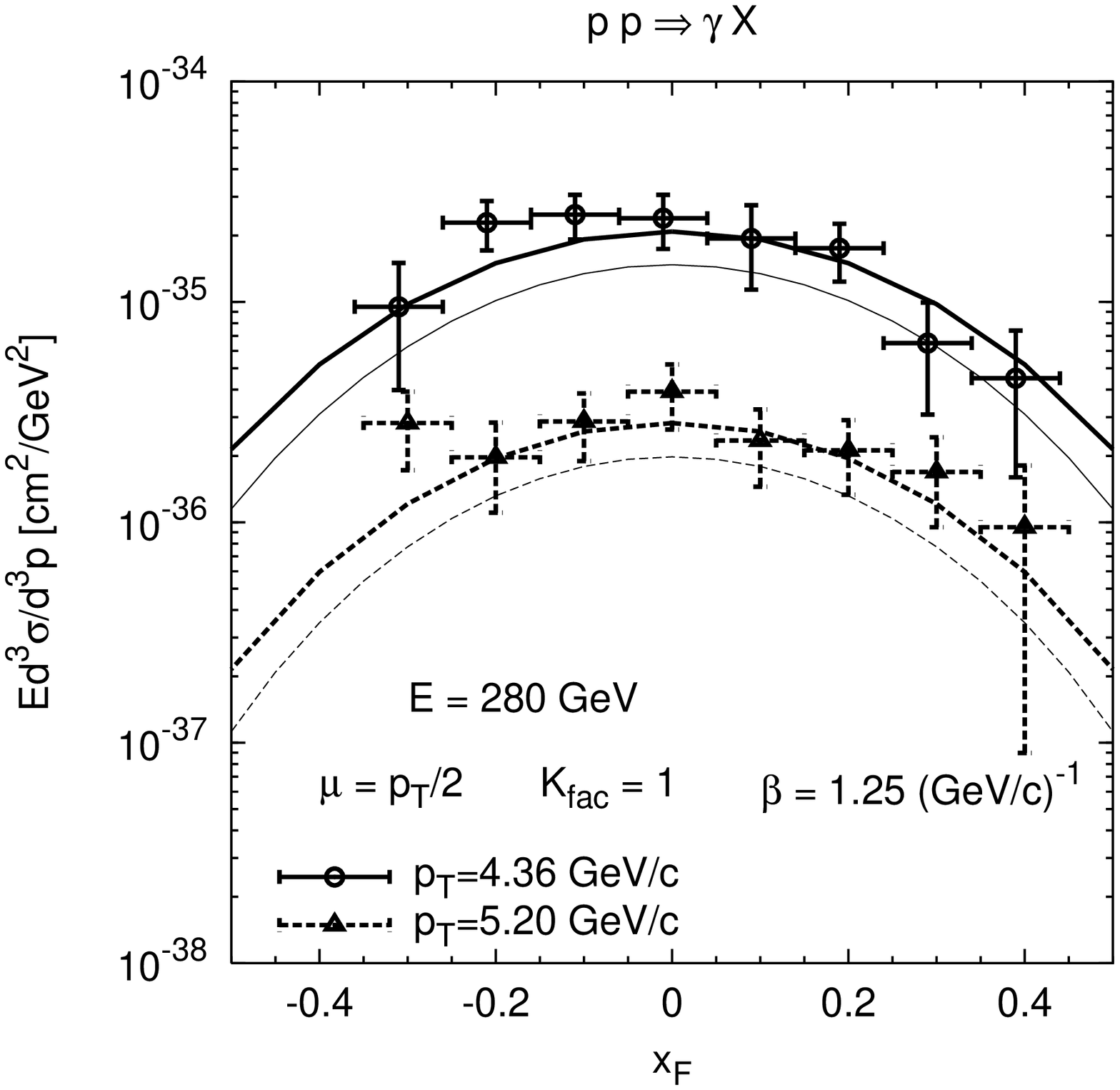}
\caption{\small 
The invariant cross section for (a) $p\,p\to\mu^+\mu^-\,X$
vs. $q_{_T}$ and (b) $p\,p\to\gamma\,X$ vs. $x_{_F}$; see plots and text for
more details.}
\end{figure}

\noindent {\it iii)}
For inclusive pion production, $p\,p\to\pi\,X$,
some experimental results for SSA are also available, and we
can see how our approach works for SSA and unpolarized cross sections
at the same time. This case is however more intricate, since we can
have $\bfk_\perp$ effects in the fragmentation process also.
The $z$ and $\bfk_\perp$ dependences in the FF are chosen according
to Eq. (\ref{gk}); a direct $z$ dependence of the $\beta'$ parameter
seems to be favored,
$1/\beta'(z)=\langle\,k_{\perp \pi}^{\,2}(z)\,\rangle^{1/2}=
1.4\,z^{1.3}\,(1-z)^{0.2}$ GeV/$c$.

Unpolarized FF are presently known with much less accuracy than
nucleon PDF. In particular, all available sets of parameterizations 
for the pion FF are for neutral pions
(or for the sum of charged pions), since $e^+e^-$ data do not allow
to separate among $\pi^+$ and $\pi^-$ case;
this can be made under further assumptions, which remain to be tested.
In Fig. 2a we present estimates of the invariant
cross section for the process $p\,p\to\pi\,X$ at $E$= 200 GeV 
vs. $x_{_F}$ for different $p_{_T}$ values.
We use two sets of FF from Kretzer (K, thin lines) \cite{kre} and
 Kniehl, Kramer, and P\"otter (KKP, thick lines) \cite{kkp},
$K_{\rm fac}=2.4$(K), 1.9(KKP), $\beta=1.25$ (GeV$/c)^{-1}$.
Data are from \cite{dona}.

Let us now consider the SSA in $p^\uparrow\,p\to\pi\,X$, within the same
approach and assuming it is generated by the Sivers effect alone,
that is from a spin and $\bfk_\perp$ effect in the PDF
inside the initial polarized proton, described by the Sivers function
$\Delta^{\!N}\!f_{q/p^\uparrow}(x,\bfk_\perp)$. There are other possible
sources for SSA, and notably the so-called Collins effect, concerning
the fragmentation of a polarized parton into the final observed pion.
These effects are not considered here.
Analogous studies have already been performed \cite{abm},
using an effective averaging on
$\bfk_\perp$ and a simplified partonic kinematics. Here we show
the first results with full treatment of $\bfk_\perp$ effects
and partonic kinematics. These results are in good qualitative
agreement with previous work. 

\begin{figure}
\includegraphics[angle=-90,width = 7.7cm]{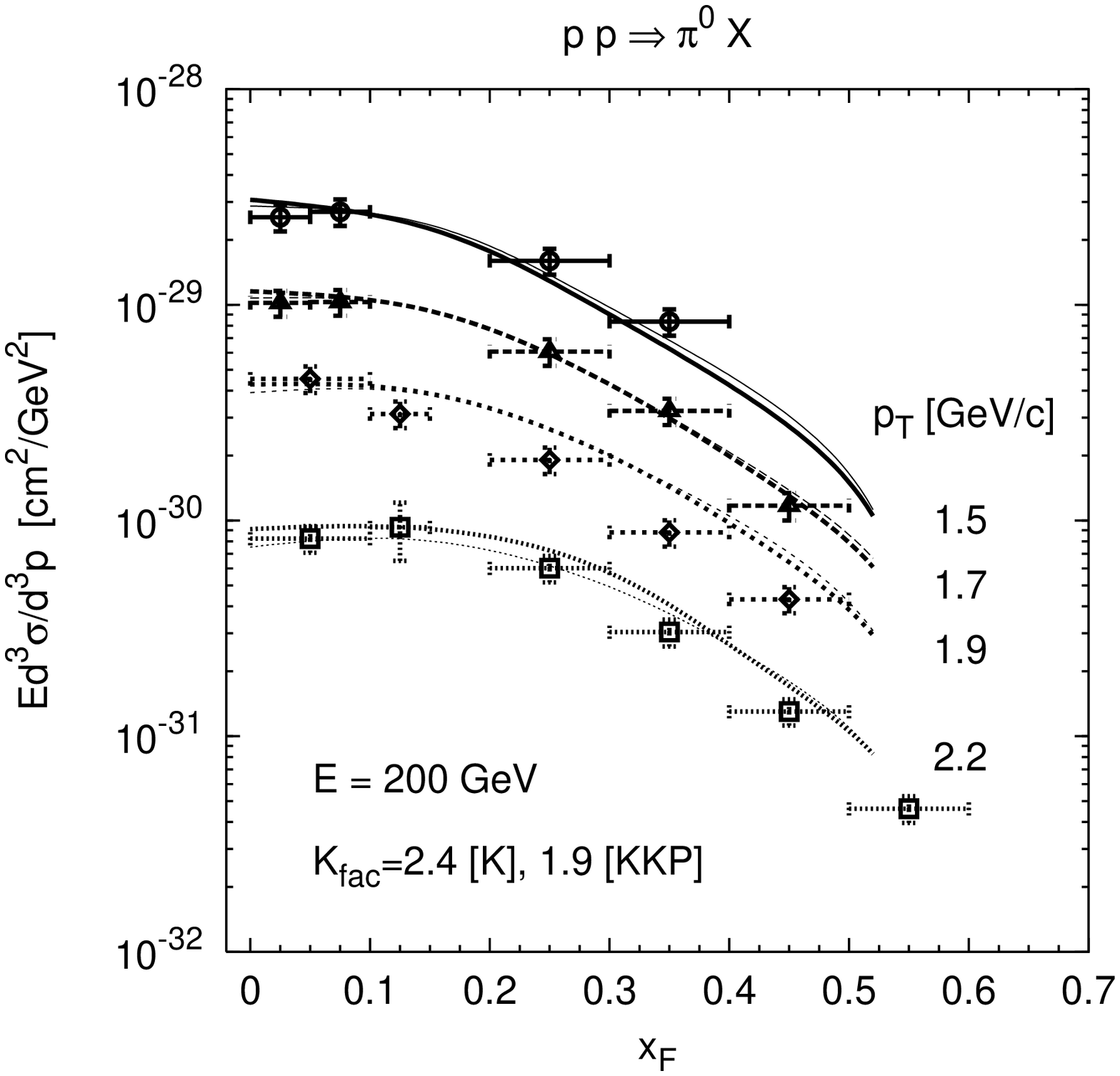}
\hspace{4pt}
\includegraphics[angle=-90,width = 7.7cm]{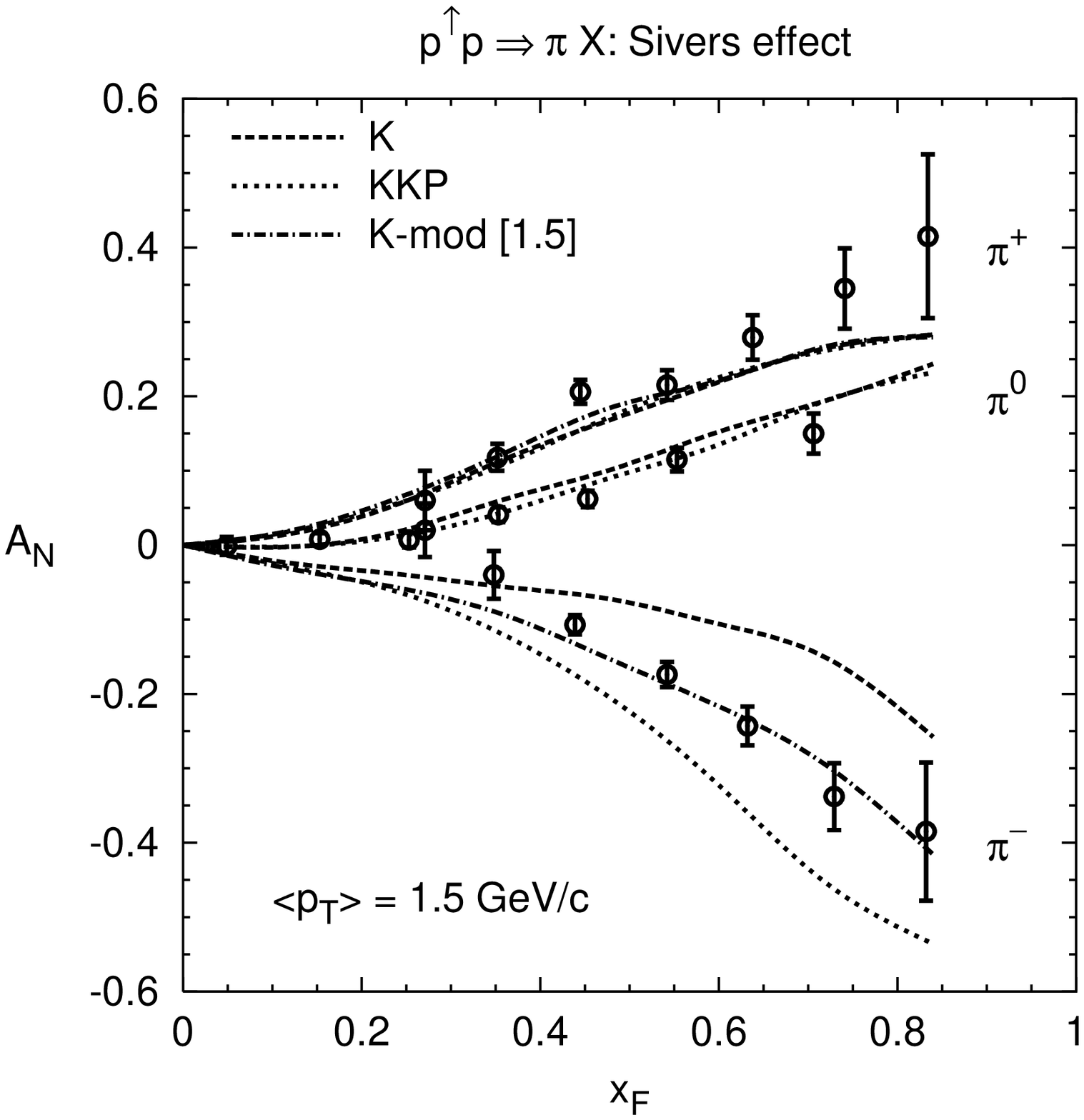}
\caption{\small
 The invariant unpol. cross section (a) and the SSA (b)
for $p^\uparrow\,p\to\pi\,X$ vs. $x_{_F}$; see plots and text for more
details.} 
\end{figure}

The numerator of the SSA, $d\sigma^\uparrow-d\sigma^\downarrow$ can be
expressed in the form of Eq.~(\ref{ucr}), with the substitution
$\hat f_{a/A}(x,\bfk_\perp)\,\to\,
\Delta^{\!N}\!f_{q/p^\uparrow}(x,\bfk_\perp)$. 
For the Sivers function we choose an expression similar to that of the
unpolarized distribution, Eq.~(\ref{gk})
\begin{equation}
\Delta^{\!N}\!f_{q/p^\uparrow}(x,\bfk_\perp)=
\Delta^{\!N}\!f_{q/p^\uparrow}(x)\,h(k_\perp)\,\sin\phi_{k_\perp}\>,
\label{dfh}
\end{equation}
\noindent where $\phi_{k_\perp}\!$ is the angle between $\bfk_\perp$ and the
polarization vector of the proton; $\Delta^{\!N}\!f_{q/p^\uparrow}(x)$
and $h(k_\perp)$ are such to fulfill the general 
positivity bound 
$|\Delta^{\!N}\!f_{q/p^\uparrow}(x,k_\perp)|/
2\,\hat f_{q/p}(x,k_\perp)$ $\leq 1$:
\begin{equation}
\Delta^{\!N}\!f_{q/p^\uparrow}(x) = N_q\,x^{a_q}(1-x)^{b_q}\,
\frac{(a_q+b_q)^{(a_q+b_q)}}{a_q^{a_q}\,b_q^{b_q}}\,
2\,f_{q/p}(x)\,,\quad
|N_q|\leq 1\,
\label{fx}
\end{equation}
\begin{equation}
h(k_\perp)=\left(2\,e\,\frac{1-r}
{r}\right)^{1/2}\,\frac{\beta^3}{\pi}\,
k_\perp\,\exp\left[\,-\beta^2k_\perp^2/r\,\right]\,
, \quad 0 < r  < 1\>.
\label{hk}
\end{equation}
A choice of the parameters in Eq.s (\ref{fx}),(\ref{hk})
which allow to reasonably reproduce the experimental results
for the pion SSA is the following (only valence quark contributions
to the Sivers function are considered):
\begin{eqnarray}
N_u &=& +0.5\quad a_u=2.0\quad b_u=0.3 \label{par}\\
N_d &=& -1.0\quad a_d=1.5\quad b_d=0.2\,,
\quad\quad\quad r \simeq 0.7\>.\nonumber
\end{eqnarray}
In Fig. 2b we show our preliminary estimates of $A_N$ with Sivers effect
at $E$ = 200 GeV and $p_{_T}$= 1.5 GeV/c, vs. $x_{_F}$, 
for three different choices of the pion FF:
K, KKP and a modified version of K. Data are from \cite{e704}.
The SSA for $\pi^+$ and $\pi^0$ is well reproduced independently of the
FF set.  
Interestingly, the $\pi^-$ case shows a stronger
sensitivity to the relation between the leading and non-leading contributions
to the fragmentation process, which cannot be extracted from present
experimental information on unpolarized pion cross sections.
In fact, our results with the K(KKP) FF sets underestimate
(overestimate) in magnitude the
$\pi^-$ asymmetry, while a good agreement is recovered using a somehow
fictitious set (K-mod) with an intermediate behavior.

In conclusion, we have presented here preliminary results of an
ongoing program dedicated to the study of partonic transverse momentum effects
both in unpolarized and polarized cross sections (and SSA) for inclusive
particle production in hadronic collisions.
These results show that it seems possible to reproduce
reasonably well, within pQCD at LO and leading twist
and up to a factor of 2-3,
unpolarized cross sections for Drell-Yan processes,
prompt photon and inclusive pion production in hadronic collisions,
in several different kinematical situations.
Within the same approach, we have reanalyzed the
SSA for $p^\uparrow\,p\to\pi\,X$ taking into account
Sivers effect alone; we
have found reasonable agreement with data and
with previous theoretical results obtained with a simplified treatment
of $\bfk_\perp$ effects and partonic kinematics, whose main results
are therefore confirmed by our analysis.
The next steps of this program are the study of the pion SSA with
Collins effect, of the SSA in photon production,
and of the unpolarized cross section and the
transverse polarization for $\Lambda$ production in
unpolarized hadronic collisions.
The extension of our analysis to RHIC kinematics,
where a thorough program on SSA measurements is in progress, 
is of great interest. First estimates of the SSA in
$p^\uparrow\,p\to\pi\,X$ seem to be in reasonable agreement with
preliminary results from RHIC \cite{rak}.

\vspace*{0.5cm}

\noindent
{\bf Acknowledgments} 

\vspace*{0.2cm}

One of us (U.D.) would like to thank the organizers for their kind
invitation to a fruitful and interesting Symposium. 
We thank Mauro Anselmino for useful discussions.   
Partial support from COFINANZIAMENTO MURST-PRIN is acknowledged. 

{\small

}

\end{document}